\documentclass[aps,prd,twocolumn,epsf,showpacs,superscriptaddress]{revtex4}
\usepackage{graphicx}% Include figure files
\usepackage{amsmath,amssymb,latexsym}
\usepackage{bm}% bold math
\usepackage{epsfig}

\newcommand{\postscript}[2]{\setlength{\epsfxsize}{#2\hsize}
   \centerline{\epsfbox{#1}}}

\def\be{\begin{equation}}
\def\ee{\end{equation}}

\def\lbldef#1#2{\expandafter\gdef\csname #1\endcsname {#2}}

\def\href#1#2{#2}

\def\ed{\end{document}}
\def\lu{\Lambda_{\cal U}}
\def\duv{d_{\rm UV}}

\newcommand{\bwide}{\begin{widetext}}
\newcommand{\ewide}{\end{widetext}}
\newcommand{\beq}[1]{\begin{equation} \label{(#1)}}
\newcommand{\eeq}{\end{equation}}
\newcommand{\ba}[1]{\begin{eqnarray} \label{(#1)}}
\newcommand{\ea}{\end{eqnarray}}

\begin{document}
%\hspace*{130mm}{\large \tt FERMILAB-PUB-07-xxx-A}

%\renewcommand{\thefootnote}{\fnsymbol{footnote}}
%\setcounter{footnote}{0}

\title{Constraints on Unparticle Physics from Solar and KamLAND Neutrinos}

\author{Luis Anchordoqui}
\affiliation{Department of Physics, University of Wisconsin-Milwaukee,
             P O Box 413, Milwaukee, WI 53201, USA}
\author{Haim Goldberg}
\affiliation{Department of Physics,
Northeastern University, Boston, MA 02115, USA}

%\date{October 2007}
\begin{abstract}
  \noindent Interest has been directed recently towards low energy
  implications of a non-trivial conformal sector of an effective field
  theory with an IR fixed point ($\Lambda_{\cal U})$, manifest in
  terms of ``unparticles'' with bizarre properties. We re-examine the
  implications of the limits on decay lifetimes of solar neutrinos for
  unparticle interactions. We study in detail the fundamental
  parameter space ($\Lambda_{\cal U}, M$) and derive bounds on the
  energy scale $M$ characterizing the new physics. We work strictly
  within the framework where conformal invariance holds down to low
  energies. We first assume that couplings of the unparticle sector to
  the Higgs field are suppressed and derive bounds with $\Lambda_{\cal
    U}$ in the TeV region from neutrino decay into scalar unparticles.
  These bounds are significant for values of the anomalous dimension
  of the unparticle operator $1.0 < d \alt 1.2$. For a region of the
  parameter space, we show that the bounds are comparable to those
  arising from production rates at high energy colliders. We then
  relax our assumption, by considering a more natural framework which
  does not require {\em a priori} restrictions on couplings of Higgs-unparticle
  operators, and derive bounds with $\Lambda_{\cal U}$ in meV region
  from neutrino decay into vector unparticles. Such low scales for the
  IR fixed point are relevant in gauge theories with many flavors.
\end{abstract}


\maketitle

A conformal hidden sector, which couples to the various gauge and
matter fields of the Standard Model (SM), has been advocated by
Georgi~\cite{Georgi:2007ek}.  In the
ultraviolet theory, the hidden sector couples to the SM through
non-renormalizable interactions
\begin{equation}
{\cal L}_{\rm UV} = \frac{O_{\rm UV}\,\, O_{\rm SM}}{M^{d_{\rm UV}+n-4}},
\end{equation}
where $M$ is the mass of the heavy exchanged particle, and $O_{\rm
  UV}$ and $O_{\rm SM}$ are hidden sector and SM operators with mass
dimensions $d_{\rm UV}$ and $n$, respectively. The hidden sector has a
non-trivial IR fixed point, $\Lambda_{\cal U}$, below which
the sector exhibits scale invariance and the operator $O_{\rm UV}$
mutates into an ``unparticle'' operator $O_{\cal U}$ with
non-integral scaling dimension $d$.  The couplings then become
\begin{equation} {\cal L}_{\cal U} = C_{\cal U} \,
 \frac{\Lambda^{d_{\rm UV}-d}}
{M^{d_{\rm UV} + n -4}} O_{\rm SM}\,\, O_{\cal U} \,\,,
\end{equation}
where $C_{\cal U}$ is a dimensionless coupling constant.

The phenomenology of the unparticle has been explored by many
groups~\cite{Georgi:2007si} and lower bounds on $\Lambda_{\cal U}$
have already been derived by considering production rates at high
energy colliders~\cite{Bander:2007nd,Rizzo:2007xr} and unparticle
emission from the core of SN1987 A~\cite{Davoudiasl:2007jr}. If
unparticle stuff exists, it could couple to the stress tensor and
mediate a new force (ungravity) between massive particles. This would
modify the inverse square law with $r$ dependence in the range between
$1/r^{4+2\delta}\ \ (\delta >0),$ a region of the parameter space
to be probed by future submillimiter
tests of gravity~\cite{Goldberg:2007tt}.

As shown in~\cite{Zhou:2007zq}, one of the bizarre implications of the
conformal hidden sector is that neutrinos would become unstable: a
neutrino mass eigenstate $\nu_j$ can decay into a another eigenstate
$\nu_i$ via $\nu_j \to \nu_i\, {\cal U},$ where ${\cal U}$ is the
invisible unparticle. In this Letter we re-examine the impact of solar
and KamLAND neutrino data on the effective couplings between neutrinos
and unparticle operators. We derive bounds on the
relevant parameter space $(\Lambda_{\cal U}, M)$, and discuss how
these bounds compare with existing limits.

Observation of solar neutrinos suggest the disappearanece of $\nu_e$
while propagating within the Sun ($\sim 2$~s) or between the Sun and
Earth ($\sim 500$~s).  Specifically, data collected by the Sudbury
Neutrino Observatory (SNO)~\cite{Ahmed:2003kj} in conjuction with data
from SuperKamiokande (SK)~\cite{Fukuda:1998fd} show that solar
$\nu_e's$ convert to $\nu_{\mu}$ or $\nu_\tau$ with CL of more than
7$\sigma$.  On the other hand, the KamLAND
Collaboration~\cite{Araki:2004mb} has measured the flux of $\overline
\nu_e$ from distant reactors and find that $\overline{\nu}_e$'s
disappear over distances of about 180~km.  The combined analysis of
solar and KamLAND neutrinos is consistent at the 3$\sigma$ CL, with
best-fit point and $1 \sigma$ ranges: $\delta m^2_\odot =
8.2^{+0.3}_{-0.3} \times 10^{-5}~{\rm eV}^2$ and $\tan^2 \theta_\odot
= 0.39^{+0.05}_{-0.04}$~\cite{Bahcall:2004ut}.  The striking agreement
between solar and KamLAND data determines a unique solution in the
mass-mixing parameter space, dubbed the Mikheyev-Smirnov-Wolfenstein Large
Mixing Angle (LMA) solution~\cite{Mikheyev:1989dy}. This provides evidence
that solar neutrinos in the energy range $5~{\rm MeV} < E_\nu < 15$~MeV are
created as nearly pure $\nu_2$ mass eigenstates.  Moreover, for LMA
their propagation in the Sun is completely adiabatic, and hence
neutrinos emerge as pure $\nu_2$ eigenstates, where $m_2 > m_1.$ The
mass ordering of neutrinos, however, is not uniquely determined.
There are two possible mass ordering that we denote as normal and
inverted, which without any loss of generality, can be chosen as
$m_1<m_2 < m_3$ and $m_3 \ll m_1 \approx m_2$, respectively. For
simplicity here we assume that the lightest neutrino is massless and take
$m_2 \approx 9$~meV and $m_3 \approx 50$~meV for the normal, and $m_2
\approx m_1 \approx 50$~meV for the inverted hierarchy.

It has long been realized that the solar neutrino flavor ratios
predicted by the standard oscillation phenomenology can be modified if
processes such as neutrino decay occur~\cite{Bahcall:1972my}. Indeed,
since neutrinos leave the Sun in a single mass eigenstate, there is no
ambiguity concerning flavor mixes at the
source~\cite{Berezhiani:1987gf}. However, in the limit that
neutrino masses are degenerate, a daughter neutrino $\nu_i$ produced
in a hypothetical decay  will carry the full energy of the
parent neutrino $\nu_j$, and would be detected by experiments on
Earth.  Therefore, the replacement of $\nu_j$ with an active daughter
$\nu_i$ of about the same energy could camouflage the characteristics
of decay. This would be specially pertinent for $\nu_2 \to \nu_1 {\cal
  U}$, where both $\nu_2$ and $\nu_1$ have large $\nu_e$ projections.
All in all, if neutrino masses are non-degenerate, the Earth-Sun baseline
defines a $\nu_2$ lifetime limit,  $\tau/m_2 \agt 10^{-4}~{\rm s}/{\rm
  eV}$~\cite{Beacom:2002cb}, for neutrinos decaying into
invisible unparticles.

Before proceeding, we stress that for any conformal field theory the
conformal dimensions of the unparticle operators are bounded from
unitarity as $d \geq 1 + s,$ where $s$ is the spin of the
operator~\cite{Mack:1975je}.  Thus, for a rank one tensor operator
$d>2$ and for a rank two $d>3$.  Because of this, vector and higher
tensor operators are less dominant in the unparticle
scheme~\cite{Nakayama:2007qu}. 

In the mass basis, the interaction between neutrinos and scalar
unparticle operators can be written as $\lambda_\nu^{ij} \, \overline
\nu_i\, \nu_j \, O_{\cal U}/\Lambda_{\cal U}^{d-1},$ where
\begin{equation}
\lambda_\nu^{ij} = C_{\cal U} \, (\Lambda_{\cal U}/M)^{d_{\rm UV} -1}
\label{lambda}
\end{equation}
is the coupling constant. The total decay rate is
found to be~\cite{Zhou:2007zq}
\begin{equation}
\Gamma_j = A_d \, \frac{|\lambda_\nu^{ij}|^2 \,\,m_j}{16 \pi^2 d
\, (d^2 -1)} \,\,
\left(\frac{m_j^2}{\Lambda_{\cal U}^2} \right)^{d-1} \,\,,
\end{equation}
where
\begin{equation}
A_{d} \equiv \frac{16 \pi^{5/2}}{(2 \pi)^{2d}} \,
\frac{\Gamma(d+1/2)}{\Gamma(d-1) \, \Gamma(2d)} \,\, .
\end{equation}
Now, using the bound on the neutrino lifetime, we can constrain the
($|\lambda_\nu^{ij}|, \Lambda_{\cal U}$) parameter space from
\begin{equation}
 \frac{16 \, \pi^2 d (d^2 -1)}{A_d \, |\lambda_\nu^{ij}|^2\,\,m_j^2} \,
\left(\frac{\Lambda_{\cal U}^2}{m_j^2}\right)^{d -1}
> 1.5 \times 10^{11}\,\, {\rm eV}^{-2} \,\, .
\label{5}
\end{equation}
Note that these constraints hide the dimension
$d_{\rm UV}$ of the Banks-Zaks (BZ) fields~\cite{Banks:1981nn}.

\begin{figure}
 \postscript{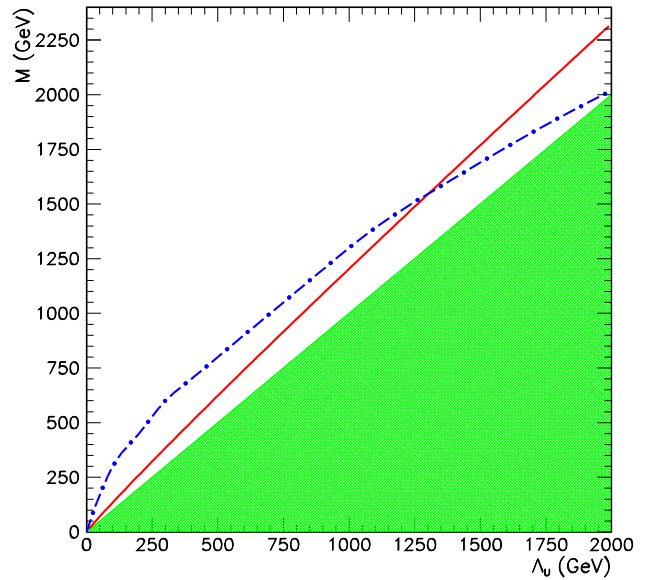}{0.98}
 \caption{Bounds from solar and KamLAND neutrinos (solid line) and
   from $e^+ e^- \to \mu^+ \mu^-$ (dot-dashed
   line)~\cite{Bander:2007nd} on the fundamental parameter space
   ($\Lambda_{\cal U},M$) for a scalar unparticle operator, with $d=
   1.1$, $d_{\rm UV} = 3,$ and couplings to the Higgs bilinear
   supressed.  The regions below the contours are excluded. The shaded
   region is excluded by the requirement $M>\Lambda_{\cal U}.$ In our
   calculations we have taken $C_{\cal U} = 0.1.$}
\label{fig:1}
\end{figure}

\begin{figure}
 \postscript{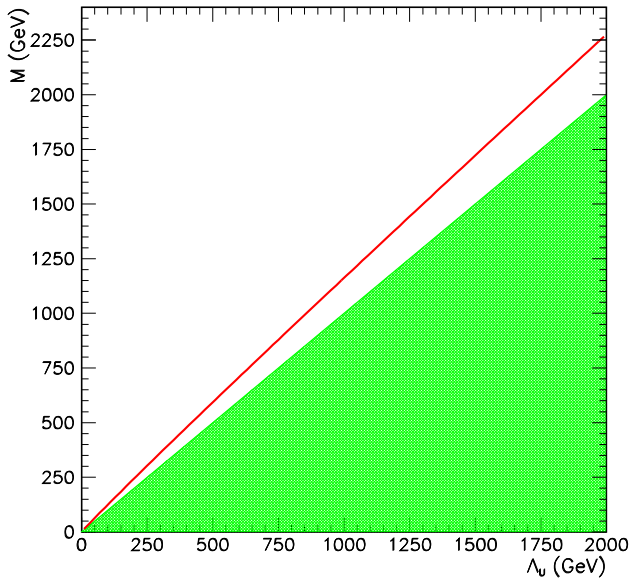}{0.98}
 \caption{Bounds from solar and KamLAND neutrinos on the fundamental
   parameter space ($\Lambda_{\cal U},M$) for a scalar unparticle
   operator, with $d= 1.1$, $d_{\rm UV} = 4,$ and couplings to the
   Higgs bilinear supressed.  The region below the contour is
   excluded. The shaded region is excluded by the requirement
   $M>\Lambda_{\cal U}.$ As in Fig.~\ref{fig:1}, we have taken
   $C_{\cal U} = 0.1.$}
\label{fig:2}
\end{figure}

To provides further probes of new physics we re-introduce the
parametrization given in Eq.~(\ref{lambda}). When combined with
Eq.~(\ref{5}), we obtain the constraint
\begin{equation}
M> D^{1/[2(d_{\rm UV}-1)]}\,
\left(\frac{\Lambda_{\cal U}}{m_j}\right)^{(1-d)/(d_{\rm UV}-1)}\
\Lambda_{\cal U}\ \ ,
\label{Mmin}
\end{equation}
where
\begin{equation}
 D=1.7\times 10^7 \ \left(\frac{m_j}{9\ {\rm meV}}\right)^2 \ 
C_{\cal U}^2 \ \frac{A_d}{16\pi^2 d(d^2-1)}\ \ .
\label{D}
\end{equation}
Since $d>1,$ the lower bound on $M$ rises faster than linearly for
small $\Lambda_{\cal U}$; then for large $\Lambda_{\cal U}$, falls
below the line $M=\Lambda_{\cal U};$ thereafter the lower bound on $M$
is simply $\Lambda_{\cal U}.$ Equating the lower bound in
Eq.~(\ref{Mmin}) to $\Lambda_{\cal U},$ we obtain the crossover
point
\begin{equation}
\Lambda_{\cal U}^{\rm cross} = D^{1/[2(d-1)]}\ m_j\ \ ,
\label{cross}
\end{equation}
independent of $\duv.$ For a qualitative view, we restrict ourselves
to the cases where the BZ~\cite{Banks:1981nn} operator is a
dimension-3 fermion bilinear $(\duv = 3)$ or a dimension-4 gauge
invariant gluon bilinear. (The dimension-3 case corresponds to the
chiral order parameter which is known to have an anomalous dimension
near $d=1$~\cite{Appelquist:1996dq}.) If it is desired to obtain a
bound in the TeV region, it is required that the crossover point occur
at $\lu\agt 1$~TeV.  A straightforward numerical exercise shows that
this occurs only when the the anomalous dimension $1<d< 1.2$ (for
$C_{\cal U}< 1).$ However, we know from the analysis
of~\cite{Fox:2007sy} that if $(a)$ the scalar operator has $d<2$ and
$(b)$ it couples to the Higgs field, then conformal invariance must be
broken at energies below the electroweak scale. This would vitiate our
bounds, because we have assumed conformal invariance down to the meV
scale. Thus, the only way to retain bounds in the TeV region is to
assume that $d\sim 1$ and that, for some reason, the coupling of the
scalar unparticle to the Higgs field vanishes.  In that case,
illustration of the results are presented in Figs.~\ref{fig:1} and
\ref{fig:2}, for the case $d = 1.1$, $C_{\cal U} = 0.1$, and $d_{\rm
  UV} = 3,\, 4$; respectively. It is evident that the results are
little changed by the variation in $d_{\rm UV}.$ The crossover point
in this case lies well above the TeV range. For comparison,
constraints on the ($\Lambda_{\cal U}, M$) parameter space from
production rate at colliders are also shown.  For $1< d \alt 1.2$,
collider bounds are comparable to those from neutrino decay into
unparticles. These bounds leave open a substantial window for
discovery of unparticle stuff at the LHC.

\begin{figure}
 \postscript{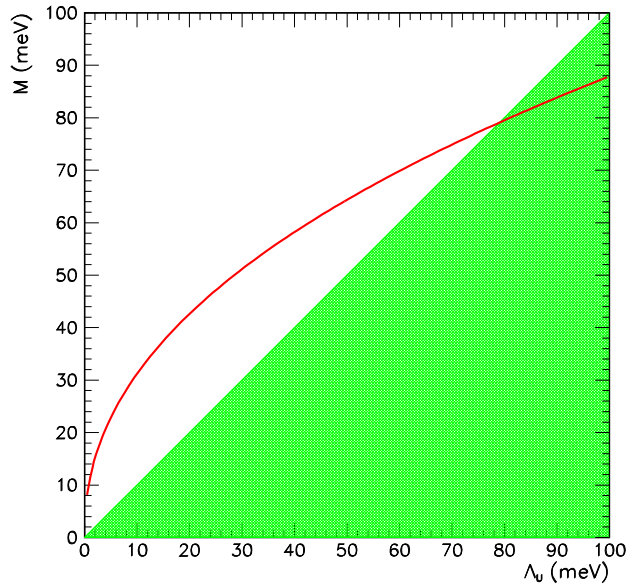}{0.98}
 \caption{Bounds from solar and KamLAND neutrinos on the fundamental
   parameter space ($\Lambda_{\cal U},M$) for a vector unparticle
   operator, with $d= 2.1$ and $d_{\rm UV} = 3$.  The region below the
   contour is excluded. The shaded region is excluded by the
   requirement $M>\Lambda_{\cal U}.$}
\label{fig:3}
\end{figure}

In what follows we consider a more natural framework, which does not
require {\em a priori} restrictions on couplings of Higgs-unparticle
operators, and we derive bounds on the ($\Lambda_{\cal U}, M$)
parameter space from neutrino decay into vector unparticles. As
mentioned above, in this case the dimension of the BZ field at the IR
fixed point runs to $d>2$.  The coupling to the Higgs field is then
naturally suppressed~\cite{Bander:2007nd}, and we can retain
conformality to low scales.  The total decay rate of neutrinos into
vector unparticles is given by~\cite{Zhou:2007zq}
\begin{equation}
\Gamma_j = 3\ A_d \, \frac{|\lambda_\nu^{ij}|^2 \,\,m_j}{16 \pi^2 d
\, (d -2) \, (d+1)} \,\,
\left(\frac{m_j^2}{\Lambda_{\cal U}^2} \right)^{d-1} \,\,,
\end{equation}
and the associated equation constraining the
($|\lambda_\nu^{ij}|, \Lambda_{\cal U}$) parameter space reads 
\begin{equation}
 \frac{16 \, \pi^2 d (d -2) (d+1)}{3 \, A_d \, 
|\lambda_\nu^{ij}|^2\,\,m_j^2} \,
\left(\frac{\Lambda_{\cal U}^2}{m_j^2}\right)^{d -1}
> 1.5 \times 10^{11}\,\, {\rm eV}^{-2} .
\end{equation}
A sample result, for $\duv = 3,\ d=2.1, \
C_{\cal U}=0.1$ is shown in Fig.~\ref{fig:3}. Such low scales for the
IR fixed point are possible in gauge theories with many
flavors~\cite{Appelquist:1996dq}.

In closing we comment on the potential of the Pierre Auger
Observatory~\cite{Abraham:2004dt} to probe the $(\Lambda_{\cal U}, M)$
plane, using cosmic baselines for measuring neutrino lifetimes.  For a
normal mass hierarchy, the relative cosmic neutrino flux
($\phi_\alpha$) on Earth would be given by the flavor ($\alpha$)
projection of the sole surviving (lightest) mass-eigenstate,
$|U_{\alpha 1}|^2$; a result that is independent of neutrino energy
and source dynamics~\cite{Beacom:2002vi}.  Because of the
$\nu_\mu$-$\nu_\tau$ interchange symmetry, each mass eigenstate
contains an equal fraction of $\nu_\mu$ and $\nu_\tau$. Unitarity plus
the condition $|U_{\mu 1}|^2 = |U_{\tau 1}|^2$ leads to earthly ratios
of $2|U_{e1}|^2 : (1 - |U_{e1}|^2): (1 - |U_{e1}|^2).$ There is then a
single flavor ratio to be determined, which we take to be
$\phi_e:\phi_\tau,$ as it can be inferred at Auger from the ratio of
measured rates of quasi-horizontal and Earth-skimming
events~\cite{Anchordoqui:2005ey}.  Substituting the measured value of
$|U_{e1}|^2$~\cite{GonzalezGarcia:2004jd}, one finds a flavor ratio
$\phi_e:\phi_\mu:\phi_\tau = 6:1:1$. This result is in striking
contrast to the expectation for stable
neutrinos~\cite{Learned:1994wg}.  Since cosmic neutrinos propagate for
distances $L \agt 100$~Mpc, future Auger observations can be used to
probe neutrino lifetimes at the level $\tau/m \sim L/E_\nu \sim
10^{-2}~{\rm s}/{\rm eV}$, increasing sensitivity to the unparticle
stuff by about half an order of magnitude.

\acknowledgments{LA is supported by UWM. HG is supported by the U.S.
  National Science Foundation (NSF) Grant No PHY-0244507.}


\begin{thebibliography}{99}


\bibitem{Georgi:2007ek}
  H.~Georgi,
  %``Unparticle Physics,''
  Phys.\ Rev.\ Lett.\  {\bf 98}, 221601 (2007)
  [arXiv:hep-ph/0703260].
  %%CITATION = PRLTA,98,221601;%%

\bibitem{Georgi:2007si}
  H.~Georgi,
  %``Another Odd Thing About Unparticle Physics,''
  Phys.\ Lett.\  B {\bf 650}, 275 (2007)
  [arXiv:0704.2457];
  %%CITATION = PHLTA,B650,275;%%
  K.~Cheung, W.~Y.~Keung and T.~C.~Yuan,
  %``Novel signals in unparticle physics,''
  arXiv:0704.2588;
  %%CITATION = ARXIV:0704.2588;%%
  M.~Luo and G.~Zhu,
  %``Some Phenomenologies of Unparticle Physics,''
  arXiv:0704.3532;
  %%CITATION = ARXIV:0704.3532;%%
  C.~H.~Chen and C.~Q.~Geng,
  %``Unparticle physics on CP violation,''
  arXiv:0705.0689;
  %%CITATION = ARXIV:0705.0689;%%
  G.~J.~Ding and M.~L.~Yan,
  %``Unparticle Physics in DIS,''
  arXiv:0705.0794;
  %%CITATION = ARXIV:0705.0794;%%
  Y.~Liao,
  %``Bounds on unparticles couplings to electrons: From electron g-2 to
  %positronium decays,''
  arXiv:0705.0837;
  %%CITATION = ARXIV:0705.0837;%%
  T.~M.~Aliev, A.~S.~Cornell and N.~Gaur,
  %``Lepton Flavour Violation in Unparticle Physics,''
  arXiv:0705.1326;
  %%CITATION = ARXIV:0705.1326;%%
  X.~Q.~Li and Z.~T.~Wei,
  %``Unparticle physics effects on D0 - anti-D0 mixing,''
  Phys.\ Lett.\  B {\bf 651}, 380 (2007)
  [arXiv:0705.1821];
  %%CITATION = PHLTA,B651,380;%%
  M.~Duraisamy,
  %``Unparticle physics in e^+ e^- annihilation,''
  arXiv:0705.2622;
  %%CITATION = ARXIV:0705.2622;%%
  C.~D.~Lu, W.~Wang and Y.~M.~Wang,
  %``Lepton flavor violating processes in unparticle physics,''
  arXiv:0705.2909;
  %%CITATION = ARXIV:0705.2909;%%
  M.~A.~Stephanov,
  %``Deconstruction of Unparticles,''
  arXiv:0705.3049;
  %%CITATION = ARXIV:0705.3049;%%
  N.~Greiner,
  %``Constraints On Unparticle Physics In Electroweak Gauge Boson Scattering,''
  arXiv:0705.3518;
  %%CITATION = ARXIV:0705.3518;%%
  D.~Choudhury, D.~K.~Ghosh and Mamta,
  %``Unparticles and Muon Decay,''
  arXiv:0705.3637;
  %%CITATION = ARXIV:0705.3637;%%
  S.~L.~Chen and X.~G.~He,
  %``Interactions of Unparticles with Standard Model Particles,''
  arXiv:0705.3946;
  %%CITATION = ARXIV:0705.3946;%%
  T.~M.~Aliev, A.~S.~Cornell and N.~Gaur,
  %``B \to K(K^*) missing energy in Unparticle physics,''
  JHEP {\bf 0707}, 072 (2007)
  [arXiv:0705.4542];
  %%CITATION = JHEPA,0707,072;%%
  P.~Mathews and V.~Ravindran,
  %``Unparticle physics at hadron collider via dilepton production,''
  arXiv:0705.4599;
  %%CITATION = ARXIV:0705.4599;%%
  G.~J.~Ding and M.~L.~Yan,
  %``Unparticle Versus NuTeV Anomaly,''
  arXiv:0706.0325;
  %%CITATION = ARXIV:0706.0325;%%
  C.~H.~Chen and C.~Q.~Geng,
  %``Unparticle phase effects,''
  arXiv:0706.0850;
  %%CITATION = ARXIV:0706.0850;%%
  Y.~Liao and J.~Y.~Liu,
  %``Long-ranged spin-spin interaction of electron from unparticle exchange,''
  arXiv:0706.1284;
  %%CITATION = ARXIV:0706.1284;%%
  K.~Cheung, W.~Y.~Keung and T.~C.~Yuan,
  %``Collider Phenomenology of Unparticle Physics,''
  arXiv:0706.3155;
  %%CITATION = ARXIV:0706.3155;%%
  R.~Zwicky,
  %``Unparticles at heavy flavour scales: CP violating phenomena,''
  arXiv:0707.0677;
  %%CITATION = ARXIV:0707.0677;%%
  T.~Kikuchi and N.~Okada,
  %``Unparticle physics and Higgs phenomenology,''
  arXiv:0707.0893;
  %%CITATION = ARXIV:0707.0893;%%
  R.~Mohanta and A.~K.~Giri,
  %``Unparticle effect on $B_s - \bar B_s$ mixing and its implications for $B_s
  %\to J/\psi \phi, \phi \phi$ decays,''
  arXiv:0707.1234;
  %%CITATION = ARXIV:0707.1234;%%
  C.~S.~Huang and X.~H.~Wu,
  %``Direct CP violation of $B \to l \nu$ in unparticle physics,''
  arXiv:0707.1268;
  %%CITATION = ARXIV:0707.1268;%%
  N.~V.~Krasnikov,
  %``Unparticle as a field with continuously distributed mass,''
  arXiv:0707.1419;
  %%CITATION = ARXIV:0707.1419;%%
  A.~Delgado, J.~R.~Espinosa and M.~Quiros,
  %``Unparticles-Higgs Interplay,''
  arXiv:0707.4309;
  %%CITATION = ARXIV:0707.4309;%%
  A.~Lenz,
  %``Unparticle physics effects in B_s mixing,''
  arXiv:0707.1535;
  %%CITATION = ARXIV:0707.1535;%%
  D.~Choudhury and D.~K.~Ghosh,
  %``Top off the unparticle,''
  arXiv:0707.2074;
  %%CITATION = ARXIV:0707.2074;%%
  H.~Zhang, C.~S.~Li and Z.~Li,
  %``Unparticle Physics and Supersymmetry Phenomenology,''
  arXiv:0707.2132;
  %%CITATION = ARXIV:0707.2132;%%
  N.~G.~Deshpande, X.~G.~He and J.~Jiang,
  %``Supersymmetric Unparticle Effects on Higgs Boson Mass and Dark Matter,''
  arXiv:0707.2959;
  %%CITATION = ARXIV:0707.2959;%%
  T.~A.~Ryttov and F.~Sannino,
  %``Conformal Windows of SU(N) Gauge Theories, Higher Dimensional
  %Representations and The Size of The Unparticle World,''
  arXiv:0707.3166;
  %%CITATION = ARXIV:0707.3166;%%
  G.~Cacciapaglia, G.~Marandella and J.~Terning,
  %``Colored Unparticles,''
  arXiv:0708.0005;
  %%CITATION = ARXIV:0708.0005;%%
  M.~Neubert,
  %``Unparticle Physics with Jets,''
  arXiv:0708.0036;
  %%CITATION = ARXIV:0708.0036;%%
  M.~x.~Y.~Luo, W.~Wu and G.~h.~Zhu,
  %``Unparticle Physics and A_{FB}^b on the Z pole,''
  arXiv:0708.0671;
  %%CITATION = ARXIV:0708.0671;%%
  N.~G.~Deshpande, S.~D.~H.~Hsu and J.~Jiang,
  %``Long range forces and limits on unparticle interactions,''
  arXiv:0708.2735;
  %%CITATION = ARXIV:0708.2735;%%
  P.~K.~Das,
  %``Unparticle effects in Supernovae cooling,''
  arXiv:0708.2812;
  %%CITATION = ARXIV:0708.2812;%%
  G.~Bhattacharyya, D.~Choudhury and D.~K.~Ghosh,
  %``Unraveling unparticles through violation of atomic parity and rare
  %beauty,''
  arXiv:0708.2835;
  %%CITATION = ARXIV:0708.2835;%%
  Y.~Liao,
  %``Impact of Unparticles on Asymptotic Freedom and Unification of Gauge
  %Couplings,''
  arXiv:0708.3327;
  %%CITATION = ARXIV:0708.3327;%%
  A.~T.~Alan and N.~K.~Pak,
  %``Unparticle pyhsics in top pair signals at the LHC,''
  arXiv:0708.3802;
  %%CITATION = ARXIV:0708.3802;%%
  I. Gogoladze, N. Okada, Q. Shafi,
  %``Unparticle physics and gauge coupling unification,''
  arXiv:0708.4405;
  %%CITATION = ARXIV:0708.4405;%%
  C.~H.~Chen and C.~Q.~Geng,
  %``Flavors and Phases in Unparticle Physics,''
  arXiv:0709.0235.
  %%CITATION = ARXIV:0709.0235;%%



\bibitem{Bander:2007nd}
  M.~Bander, J.~L.~Feng, A.~Rajaraman and Y.~Shirman,
  %``Unparticles: Scales and High Energy Probes,''
  arXiv:0706.2677.
  %%CITATION = ARXIV:0706.2677;%%

\bibitem{Rizzo:2007xr}
  T.~G.~Rizzo,
  %``Contact Interactions and Resonance-Like Physics at Present and Future
  %Colliders from Unparticles,''
  arXiv:0706.3025.
  %%CITATION = ARXIV:0706.3025;%


\bibitem{Davoudiasl:2007jr}
  H.~Davoudiasl,
  %``Constraining Unparticle Physics with Cosmology and Astrophysics,''
  arXiv:0705.3636;
  %%CITATION = ARXIV:0705.3636;%%
  S.~Hannestad, G.~Raffelt and Y.~Y.~Y.~Wong,
  %``Unparticle constraints from SN1987A,''
  arXiv:0708.1404;
  %%CITATION = ARXIV:0708.1404;%%
  A. Freitas and D. Wyler,
  %``Astro unparticle Physics'',
  arXiv:0708.4339.
  %%CITATION = ARXIV:0708.4339;%%


\bibitem{Goldberg:2007tt}
  H.~Goldberg and P.~Nath,
  %``Ungravity and Its Possible Test,''
  arXiv:0706.3898.
  %%CITATION = ARXIV:0706.3898;%%

\bibitem{Zhou:2007zq}
  S.~Zhou,
  %``Neutrino Decays and Neutrino Electron Elastic Scattering in Unparticle
  %Physics,''
  arXiv:0706.0302;
  %%CITATION = ARXIV:0706.0302;%%
%\bibitem{Chen:2007zy}
  S.~L.~Chen, X.~G.~He and H.~C.~Tsai,
  %``Constraints on Unparticle Interactions from Invisible Decays of Z,
  %Quarkonia and Neutrinos,''
  arXiv:0707.0187;
  %%CITATION = ARXIV:0707.0187;%%
  X.~Q.~Li, Y.~Liu and Z.~T.~Wei,
  %``Neutrino decay as a possible interpretation to the MiniBooNE observation
  %with unparticle scenario,''
  arXiv:0707.2285;
  %%CITATION = ARXIV:0707.2285;%%
  D.~Majumdar,
  %``Unparticle decay of neutrinos and it's effect on ultra high energy
  %neutrinos,''
  arXiv:0708.3485.
  %%CITATION = ARXIV:0708.3485;%%





\bibitem{Ahmed:2003kj}
  S.~N.~Ahmed {\it et al.}  [SNO Collaboration],
  %``Measurement of the total active B-8 solar neutrino flux at the Sudbury
  %Neutrino Observatory with enhanced neutral current sensitivity,''
  Phys.\ Rev.\ Lett.\  {\bf 92}, 181301 (2004)
  [arXiv:nucl-ex/0309004].
  %%CITATION = NUCL-EX 0309004;%%

\bibitem{Fukuda:1998fd}
  Y.~Fukuda {\it et al.}  [Super-Kamiokande Collaboration],
  %``Measurements of the solar neutrino flux from Super-Kamiokande's first  300
  %days,''
  Phys.\ Rev.\ Lett.\  {\bf 81}, 1158 (1998)
  [Erratum-ibid.\  {\bf 81}, 4279 (1998)]
  [arXiv:hep-ex/9805021];
  %%CITATION = HEP-EX 9805021;%%
%\bibitem{Fukuda:1998rq}
  Y.~Fukuda {\it et al.}  [Super-Kamiokande Collaboration],
  %``Constraints on neutrino oscillation parameters from the measurement of
  %day-night solar neutrino fluxes at Super-Kamiokande,''
  Phys.\ Rev.\ Lett.\  {\bf 82}, 1810 (1999)
  [arXiv:hep-ex/9812009];
  %%CITATION = HEP-EX 9812009;%%
%\bibitem{Fukuda:1998ua}
  Y.~Fukuda {\it et al.}  [Super-Kamiokande Collaboration],
  %``Measurement of the solar neutrino energy spectrum using neutrino  electron
  %scattering,''
  Phys.\ Rev.\ Lett.\  {\bf 82}, 2430 (1999)
  [arXiv:hep-ex/9812011];
  %%CITATION = HEP-EX 9812011;%%
%\bibitem{Fukuda:2001nj}
  S.~Fukuda {\it et al.}  [Super-Kamiokande Collaboration],
  %``Solar B-8 and he p neutrino measurements from 1258 days of
  %Super-Kamiokande data,''
  Phys.\ Rev.\ Lett.\  {\bf 86}, 5651 (2001)
  [arXiv:hep-ex/0103032];
  %%CITATION = HEP-EX 0103032;%%
%\bibitem{Fukuda:2001nk}
  S.~Fukuda {\it et al.}  [Super-Kamiokande Collaboration],
  %``Constraints on neutrino oscillations using 1258 days of  Super-Kamiokande
  %solar neutrino data,''
  Phys.\ Rev.\ Lett.\  {\bf 86}, 5656 (2001)
  [arXiv:hep-ex/0103033];
  %%CITATION = HEP-EX 0103033;%%
%\bibitem{Fukuda:2002pe}
  S.~Fukuda {\it et al.}  [Super-Kamiokande Collaboration],
  %``Determination of solar neutrino oscillation parameters using 1496 days  of
  %Super-Kamiokande-I data,''
  Phys.\ Lett.\ B {\bf 539}, 179 (2002)
  [arXiv:hep-ex/0205075].
  %%CITATION = HEP-EX 0205075;%%





\bibitem{Araki:2004mb}
  T.~Araki {\it et al.}  [KamLAND Collaboration],
  %``Measurement of neutrino oscillation with KamLAND: Evidence of spectral
  %distortion,''
  Phys.\ Rev.\ Lett.\  {\bf 94}, 081801 (2005)
  [arXiv:hep-ex/0406035].
  %%CITATION = HEP-EX 0406035;%%




\bibitem{Bahcall:2004ut}
  J.~N.~Bahcall, M.~C.~Gonzalez-Garcia and C.~Pena-Garay,
  %``Solar neutrinos before and after Neutrino 2004,''
  JHEP {\bf 0408}, 016 (2004)
  [arXiv:hep-ph/0406294].
  %%CITATION = HEP-PH 0406294;%%




\bibitem{Mikheyev:1989dy}
  S.~P.~Mikheyev and A.~Y.~Smirnov,
  %``Resonant neutrino oscillations in matter,''
  Prog.\ Part.\ Nucl.\ Phys.\  {\bf 23}, 41 (1989);
  %%CITATION = PPNPD,23,41;%%
%\bibitem{Wolfenstein:1977ue}
  L.~Wolfenstein,
  %``Neutrino oscillations in matter,''
  Phys.\ Rev.\  D {\bf 17}, 2369 (1978).
  %%CITATION = PHRVA,D17,2369;%%



\bibitem{Bahcall:1972my}
  J.~N.~Bahcall, N.~Cabibbo and A.~Yahil,
  %``Are Neutrinos Stable Particles?,''
  Phys.\ Rev.\ Lett.\  {\bf 28}, 316 (1972);
  %%CITATION = PRLTA,28,316;%%
%\bibitem{Pakvasa:1972gz}
  S.~Pakvasa and K.~Tennakone,
  %``Neutrinos of Non-Zero Rest Mass,''
  Phys.\ Rev.\ Lett.\  {\bf 28}, 1415 (1972);
  %%CITATION = PRLTA,28,1415;%%
%\bibitem{Bahcall:1986gq}
  J.~N.~Bahcall, S.~T.~Petcov, S.~Toshev and J.~W.~F.~Valle,
  %``Tests Of Neutrino Stability,''
  Phys.\ Lett.\  B {\bf 181}, 369 (1986).
  %%CITATION = PHLTA,B181,369;%%



\bibitem{Berezhiani:1987gf} When neutrinos propagate in dense matter,
  their decay rates can be increased by the greater phase space
  provided by the matter enhancement to $\delta m^2$.  [
  Z.~G.~Berezhiani and M.~I.~Vysotsky,
  %``Neutrino decay in matter,''
  Phys.\ Lett.\  B {\bf 199}, 281 (1987).]
  %%CITATION = PHLTA,B199,281;%%
  However, for LMA-type solutions this effect is small and decay in
  the Sun can be neglected.






\bibitem{Beacom:2002cb}
  J.~F.~Beacom and N.~F.~Bell,
  %``Do solar neutrinos decay?,''
  Phys.\ Rev.\  D {\bf 65}, 113009 (2002)
  [arXiv:hep-ph/0204111].
  %%CITATION = PHRVA,D65,113009;%%

\bibitem{Mack:1975je}
  G.~Mack,
  %``All Unitary Ray Representations Of The Conformal Group SU(2,2) With
  %Positive Energy,''
  Commun.\ Math.\ Phys.\  {\bf 55}, 1 (1977).
  %%CITATION = CMPHA,55,1;%%




\bibitem{Nakayama:2007qu}
  Y.~Nakayama,
  %``SUSY Unparticle and Conformal Sequestering,''
  arXiv:0707.2451.
  %%CITATION = ARXIV:0707.2451;%%

\bibitem{Banks:1981nn}
  T.~Banks and A.~Zaks,
  %``On The Phase Structure Of Vector-Like Gauge Theories With Massless
  %Fermions,''
  Nucl.\ Phys.\  B {\bf 196}, 189 (1982).
  %%CITATION = NUPHA,B196,189;%%


\bibitem{Appelquist:1996dq}
  T.~Appelquist, J.~Terning and L.~C.~R.~Wijewardhana,
  %``The Zero Temperature Chiral Phase Transition in SU(N) Gauge Theories,''
  Phys.\ Rev.\ Lett.\  {\bf 77}, 1214 (1996)
  [arXiv:hep-ph/9602385];
  %%CITATION = PRLTA,77,1214;%%
  T.~Appelquist, in Particles and Fields (Mexican School), J. L. Lucio
  and A. Zepeda eds. (World Scientific, 1992).



\bibitem{Fox:2007sy}
  P.~J.~Fox, A.~Rajaraman and Y.~Shirman,
  %``Bounds on Unparticles from the Higgs Sector,''
  arXiv:0705.3092.
  %%CITATION = ARXIV:0705.3092;%%

\bibitem{Abraham:2004dt}
  J.~Abraham {\it et al.}  [Pierre Auger Collaboration],
  %``Properties and performance of the prototype instrument for the Pierre Auger
  %Observatory,''
  Nucl.\ Instrum.\ Meth.\  A {\bf 523}, 50 (2004).
  %%CITATION = NUIMA,A523,50;%%


\bibitem{Beacom:2002vi}
  J.~F.~Beacom, N.~F.~Bell, D.~Hooper, S.~Pakvasa and T.~J.~Weiler,
  %``Decay of high-energy astrophysical neutrinos,''
  Phys.\ Rev.\ Lett.\  {\bf 90}, 181301 (2003)
  [arXiv:hep-ph/0211305].
  %%CITATION = PRLTA,90,181301;%%

\bibitem{Anchordoqui:2005ey}
  L.~Anchordoqui, T.~Han, D.~Hooper and S.~Sarkar,
  %``Exotic neutrino interactions at the Pierre Auger observatory,''
  Astropart.\ Phys.\  {\bf 25}, 14 (2006)
  [arXiv:hep-ph/0508312];
  %%CITATION = APHYE,25,14;%%
%\bibitem{Anchordoqui:2001cg}
  L.~A.~Anchordoqui, J.~L.~Feng, H.~Goldberg and A.~D.~Shapere,
  %``Black holes from cosmic rays: Probes of extra dimensions and new limits
  %on TeV-scale gravity,''
  Phys.\ Rev.\  D {\bf 65}, 124027 (2002)
  [arXiv:hep-ph/0112247].
  %%CITATION = PHRVA,D65,124027;%%

\bibitem{Learned:1994wg}
  J.~G.~Learned and S.~Pakvasa,
  %``Detecting tau-neutrino oscillations at PeV energies,''
  Astropart.\ Phys.\  {\bf 3}, 267 (1995)
  [arXiv:hep-ph/9405296];
  %%CITATION = APHYE,3,267;%%
  The predicted difference in flavor ratios for stable and unstable
  neutrinos remains statistically significantly away from 6:1:1 when
  considering uncertainties in source dynamics.  P.~Lipari,
  M.~Lusignoli and D.~Meloni,
  %``Flavor Composition and Energy Spectrum of Astrophysical Neutrinos,''
  Phys.\ Rev.\  D {\bf 75}, 123005 (2007)
  [arXiv:0704.0718].
  %%CITATION = PHRVA,D75,123005;%%

\bibitem{GonzalezGarcia:2004jd}
  M.~C.~Gonzalez-Garcia,
  %``Global analysis of neutrino data,''
  Phys.\ Scripta {\bf T121}, 72 (2005)
  [arXiv:hep-ph/0410030].
  %%CITATION = PHSTB,T121,72;%%



\end{thebibliography}
\end{document}